# Chapter 6
# Economic Origins of the Sicilian Mafia: A Simulation Feedback Model[*]

Oleg V. Pavlov
Worcester Polytechnic Institute
Worcester, MA   USA
opavlov@wpi.edu

Jason M. Sardell
The University of Texas at Austin
Austin, TX   USA
jsardell@austin.utexas.edu

## 6.1  Introduction

During the nineteenth century, Sicily underwent a series of political and land reforms that led to dismantling of large land ownership and culminated in the unification of the Italian state in 1861. In the new economic and political regime, banditry and a general sense of lawlessness proliferated (Del Monte and Pennacchio 2012). The Italian government proved ineffective at enforcing the rule of law on the island (Buonanno et al. 2015; Dimico et al. 2017; Acemoglu et al. 2020). In that environment, the mafia emerged as a private response to the inadequate protection by the state (Gambetta 1993). The mafia offered private protection from bandits to landowners who could not rely on protection and enforcement of property rights by the state (Blok 1969; Bandiera 2003).

In many parts of the world, organized crime has acted as a parallel legal system, taking advantage of the high demand for protection and enforcement services during periods characterized by the absence of effective government institutions (Blok 1969; Skaperdas 2001). Research has documented similarities between organized crime syndicates in Sicily, Japan and Russia (Gambetta 1993; Bandiera 2003). As in Sicily, yakuza emerged in post-feudal Japan (Bandiera 2003; Milhaupt and West 2000). Based on extensive data, Milhaupt and West (2000) show that yakuza complemented inefficiencies in state legal and enforcement structures by providing contract enforcement and private property protection. Russian mafia emerged during the post-Communist transition of the 1990s when legal institutions required for the market economy were weak (Varese 1994). In interviews, small businesses in Russia indicated that they viewed mafia as a substitute to police protection and state-provided courts (Frye and Zhuravskaya 2001).

This chapter builds a *feedback-rich theory* of economic origins of the Sicilian mafia. The study contributes to the active research on economic causes of organized crime (e.g., Skaperdas 2001; Anderson and Bandiera 2006; Buonanno et al. 2015; Dimico et al. 2017; Acemoglu et al. 2020). We develop a *simulation feedback model* that is consistent with earlier empirical and theoretical research on organized crime. Our model captures the causal links between the mafia's activity, weak law enforcement, and profitability of businesses in the region. Organized crime is seen as the enforcer of private property when the rule of law is weak. The model offers a theoretical framework that explains the emergence of the demand for private protection and explains variations in mafia activity across regions. It shows that weak law enforcement and valuable loot-prone production can foster mafia emergence and persistence.

Rooted in the system dynamics methodology, feedback models allow detailed analysis in and out of equilibrium. System dynamics (e.g., Sterman 2000; Maani and Cavana 2007; Morecroft 2007; Galbraith 2010; Duggan 2018) offers a set of tools for construction of comprehensive theories that combine complex causal connections between variables. Feedback models allow us to explore and observe the change mechanisms and the workings of distinct effects as they pull on variables in different directions.

The chapter is organized as follows. The next section provides the historical context of the situation in Sicily at the end of the nineteenth century. Section 6.3 reviews models of criminal behavior. The system dynamics method is briefly explained in Section 6.4. Section 6.5 describes the model, followed by the analysis of its key feedback loops in Section 6.6. Simulation experiments are presented in Section 6.7. Final comments and suggestions for future research are provided in Section 6.8.

## 6.2   Historical context

Prior to the nineteenth century, land in Sicily was owned by few noble families (Blok 1966; Bandiera 2003). Feudal landowners also controlled all aspects of life including tax collection and administration of justice (Dimico et al. 2017). Because the land-owning aristocrats preferred to live in Naples or Palermo, many of them either hired *soprastante* to oversee the lands or leased their property to *gabellotti* who managed the lands in their stead (Blok 1966; 1969). *Soprastante* and *gabelllotti*, in turn, recruited *campieri,* or soldiers, into small private armies that protected the estates and also policed the peasants (Blok 1966; Bandiera 2003; Dimico et al. 2017).

A series of reforms between 1812 and 1863 abolished feudalism, introduced private property, confiscated common and church lands, and redistributed lands from nobility to new landowners (Bandiera 2003; Del Monte and Pennacchio 2012). As a result, the number of landowners in Sicily increased from 2,000 to 20,000 (Blok 1969; Bandiera 2003). The vast majority of these properties were sold to *gabellotti* (Blok 1966; 1969). Those lands that were sold to peasants tended to be very small and of the lowest quality. Thus, peasants often earned their livelihood by working as sharecroppers for large landowners (Blok 1966).

Sharecroppers' earnings were meager (Blok 1966). Peasants also lost rights to common lands without any compensation (Del Monte and Pennacchio 2012). For example, confiscation of church lands and the sale of these large estates to private hands meant that peasants no longer had access to traditional sources of supplementary income such as firewood gathering and livestock grazing (Blok 1969). To make ends meet, some peasants turned to crime (Blok 1966; Blok 1969; Bandiera 2003). The land-reform era of 1812-1863 saw a rise of banditry in Sicily (Bandiera 2003; Dimico et al. 2017), especially in poorer areas (Del Monte and Pennacchio 2012).

State-provided public security was ineffective as the police was severely understaffed (Bandiera 2003). There were usually fewer than 350 policemen stationed in the entire island, or one per 28 square miles (Bandiera 2003). A Sicilian magistrate noted in 1874 that "people in the countryside [are] more afraid of criminals than they are of the Law … [the] public force is completely overcome by the strength of the criminals" (Bandiera 2003). Additionally, many policemen were former bandits and criminals, which led to rampant collusion with bandits (Bandiera 2003). The trust of the law was so low that it was considered a civic duty to keep the police out of local affairs in some communities (Brögger 1968).

The abolition of feudalism and the creation of the Italian state thus resulted in a power vacuum that was conducive for the emergence of the mafia (Gambetta 1993). With the private armies dismantled and the new state incapable or unwilling to provide protection and enforcement, the mafia stepped in to offer private protection services to landowners (Gambetta 1993; Dimico et al. 2017). Indeed, banditry was less prevalent in areas where mafia was active (Del Monte and Pennacchio 2012). However, the mafia was not uniformly active across Sicily. In the mid-1880s, mafia activity existed in about half of the areas populated mostly by medium and small landowners; it was significantly less active in areas with large landholdings (Bandiera 2003). High economic output combined with weak law and order created demand for private protection (Del Monte and Pennacchio 2012; Buonanno et al. 2015; Dimico et al. 2017).

Analyzing original court data from 1883, Dimico et al. (2017) connected the emergence of the mafia to areas with intensive farming of oranges and lemons (Dimico et al. 2017). They argue that an exogenous demand shock for citrus fruits created incentives for the emergence of the Sicilian mafia (Dimico et al. 2017). Lemon farms were about 30 times more profitable than grapes or olives, but were more difficult to secure from theft (Dimico et al. 2017). In high-profit areas with citrus farming, the mafia protected the crops from criminals (Dimico et al. 2017). Buonanno et al. (2015) found that presence of sulphur mines was a strong predictor of mafia activity round 1900s. Sulphur was important for the economy of Sicily through the entire nineteenth century – it was the largest export in terms of the Sicilian GDP (Buonanno et al. 2015).

Bandits and mafia had a symbiotic relationship. Mafiosi were most often recruited from former bandits, especially the ones who had a proven record of success and violence, as their reputation made them particularly suited to the requirements of the job (Blok 1969; Bandiera 2003). And the mafia needed the bandits to justify its services (Buonanno et al. 2015). The mafia used the threat of uncontrolled banditry to force their protection services on the landlords and peasants. Moreover, the mafia recognized the source of their power and would occasionally allow bandits to operate with impunity in exchange for a share of the profits (Blok 1966). For example, Blok (1966) noted that cattle-rustling was a common problem, although it would have been impossible for local bandits to steal cattle without the assistance of the local mafia. Instead of putting bandits out of business, the mafia often assisted bandits in crimes such as cattle-rustling (Blok 1966). This way, the mafia earned additional income from illegal activities as a tribute paid by bandits (Blok 1966).

## 6.3   Models of criminal behavior

This section reviews earlier literature on theoretical models of criminal behavior that is typically divided into banditry and organized crime, or mafia. These models are often variants of a three-class theoretical framework, which is relatively simple yet convenient for describing criminal activities. These models include a productive class called farmers, peasants, or traders. They are predated on by bandits. The output of the productive class is protected by the enforcer class referred to as soldiers, rulers, cops, or mafia. The cost of the protection is paid by the productive class.

Usher (1989) uses a three-class analytical framework that consists of the interacting populations of rulers, farmers, and bandits. The growth of the peasant population is linked to the economic conditions and casualties that depend on the extent of banditry. Usher's model generates oscillations between anarchy and despotism. The farmers and the bandits are dominant populations in the anarchy state, while rulers are in control during the despotism phase.

Feichtinger and Novak (1994) develop a three-class model that they analyze using the game-theoretic approach. In their framework, farmers generate wealth that the bandits plunder. The bandit loot depends on the farmer wealth and the success of the plundering effort. Rulers attempt to prevent bandits from plundering and in exchange for protection, rulers levy taxes on farmers. Bandits and rulers choose their effort levels by optimizing their respective inter-temporal net utilities. The model generates a stable limit cycle of increasing and waning dominance of bandits or rulers.

Feichtinger et al. (1996) propose a predator-prey model populated by distinct and non-mixing groups of rulers, soldiers and farmers. Farmers are prey exploited by bandits and soldiers. The population of bandits is controlled by the soldiers who tax farmers for protection services. The mortality of the farmer population is a sum of the direct deaths from bandits and indirect deaths due to the taxation by soldiers. The bandit population swells in the years when the loot is plentiful, but it is reduced through natural deaths and the casualties inflicted by soldiers.

Bandiera (2003) develops a two-stage game-theoretical model to analyze the system that led to the rise of the mafia in Sicily. In the first stage, landlords offer a payoff to the mafia for property protection. In the second stage, the mafia decides how much land to protect given the offer while maximizing its revenue. Bandits are not explicitly modeled as part of the game even though they are in the model indirectly as thieves who steal income from landlords. In this framework, landlords buying protection impose a negative externality on the landlords who do not buy protection because unprotected landlords are more likely to be robbed. The model predicts that the negative externality forces all rational landowners to purchase protection services. It also shows that the mafia has an incentive to allow some level of robberies in order to stimulate demand for protective services. The model predicts that greater property fragmentation and greater land income stimulate the demand for protection.

Anderson and Bandiera (2006) develop a model of trade that includes predation by robbers and enforcement by cops. Traders costlessly acquire goods that they transport, and robbers steal from traders. Robbers are recruited from the ranks of traders. Two states are analyzed – anarchy and private enforcement. In anarchy, traders have no protection. In the other arrangement, private cops protect traders for a fee, which reduces the probability of success for robbers. The authors note that the mafia is equivalent to private cops operating as a protection monopoly. Their model differentiates between the true and perceived probabilities of trader's ability to avoid being robbed. It does not endogenize the mafia labor supply. The model shows that both trade and predation coexist for certain parameter values. Trade may increase or decrease when enforcement in present, and eliminating the mafia has an ambiguous effect on trade.

Saeed and Pavlov (2008) build a simulation feedback model that includes populations of farmers, bandits and soldiers. Farmers produce, bandits predate on farmers, and soldiers protect farmers from bandits. Bandits and soldiers are recruited from the population of farmers. Bandits may choose to resume farming, and soldiers retire to farming when fewer soldiers are needed. Protection by soldiers is costly and it is paid through taxation. The model generates cycles of anarchy and tyranny before converging to equilibrium.

Del Monte and Pennacchio (2012) develop a model to explain why after Italian unification, mafia activity in Southern Italy was greater in regions with high output and weak rule of law. To test the model, they compare historic data on regions with high and low incomes. Assuming a fixed cost of protection, the model shows that the market for protection exists only when the expected loss from crime is greater than some threshold and when the willingness to pay for protection is greater than the cost. The model also shows that mafia is typically less active, and sometimes not active at all, in cases of low output, even if banditry is common.

Saeed et al. (2013) expand the simulation three-class framework in Saeed and Pavlov (2008) to include psychological factors. Recruitment into population groups of farmers, bandits and soldiers is affected by economic incentives and by psychological factors such as group identity and exposure to violence. For example, being a farmer is less attractive when farmers are subjected to high levels of violence by bandits. Psychological effects expedite adjustments to equilibria and provide additional policy space. Experiments suggest that higher levels of output and greater respect for law and order lead to fewer crimes, and therefore fewer soldiers.

Konrad and Skaperdas (2012) analyze competition in a private market for protection. They employ a framework of honest peasants, predatory bandits, and warriors who are the protectors of peasants and who are paid by the peasants for protection services. The authors study two market structures: monopolistic competition and monopoly. Because mafia groups compete using force, more competition between violent protectors leads to inferior outcomes for peasants as compared to buying protection from a monopolistic protector or not buying protection at all. The authors also consider a case in which peasants self-organize collective protection against bandits, but find that such a state is not stable, and therefore not sustainable.

## 6.4  System dynamics method

This chapter uses system dynamics to model interactions between peasants, bandits, and the mafia. The fundamental premise of system dynamics is that structure causes behavior. Rooted in control engineering, systems dynamics was invented in the late 1950s as a tool to study management problems (Forrester 1958; 1961). It was later applied to urban management (Forrester 1969). The method has become the second most widely used simulation approach in operations research (Jahangirian et al. 2010). Applications include management (for example, Davis et al. 2007; Maani and Cavana 2007), strategy (e.g., Warren 2002; Morecroft 2007), information technology (e.g., Pavlov and Saeed 2004; Georgantzas and Katsamakas 2008), health (e.g., Pedercini et al. 2011; Zainal et al. 2014), and political science (e.g., Wils et al. 1998; Pavlov et al. 2005; Langarudi and Radzicki 2018). Universities around the world offer courses and academic programs in systems dynamics (Davidsen et al. 2014; Pavlov et al. 2014; Schaffernicht and Groesser 2016; Cavana and Forgie 2018).

The system dynamics approach (Sterman 2000; Maani and Cavana 2007; Morecroft 2007; Galbraith 2010; Duggan 2018) includes identifying complex causal chains that form feedback loops and cause delays that are translated into a computational model. The model boundary and the simulated time interval are dependent on the studied problem. The causal structure of a system dynamics model is communicated using relatively simple graphic notation (Lane 2000) that is reminiscent of signed directed graphs. Variables that accumulate over time are called stocks and denoted in diagrams as rectangles. Stocks increase through inflows and decline through outflows, which are drawn as miniature pipes with faucets. Inflows and outflows are controlled by other variables in the model. Mathematically, a system dynamics model is a set of nonlinear non-stochastic partial differential equations that are solved numerically. There are many software packages that are tailored for system dynamics modeling. This chapter uses *Stella Architect* from the company called *isee systems*[1].

## 6.5  Model development

This section develops the *simulation feedback model*[2]. Its formulation follows closely the theoretical literature reviewed earlier. The model accounts only for economic reasons that encourage or suppress banditry and mafia existence. Violence, imprisonment, social norms or psychological factors are left outside the boundaries of the model. In this model, mafia activity discourages banditry because it makes banditry less attractive economically, rather than through threats of violence.

As in Bandiera (2003), Saeed and Pavlov (2008) and Saeed et al. (2013), we assume constant population size. The population consists of three groups: peasants, $P$, bandits, $B$, and the mafia, $M$. Peasants are honest producers who are engaged in constant returns to scale production. The total peasant output is $Y = a_P P$, where $a_P > 0$ is the marginal product and $P$ is the peasant population. In the spirit of Bandiera (2003), we assume that the marginal cost of production is zero.

---

[1] https://www.iseesystems.com/ . Last accessed on June 24, 2020.
[2] An electronic version of the model is available as an online supplementary file with this chapter.

As in Bandiera (2003) and Anderson and Bandiera (2006), landowners who buy protection are not 100 percent immune from theft and a portion of the total peasant output is lost to bandits. Following Anderson and Bandiera (2006), the probability of avoiding banditry is represented by the function $F(B/P) = 1/(1 + \theta_B B/P)$, where parameter $\theta_B > 0$ describes the effectiveness of bandit technology. Function $F(B/P) \in [0,1]$ is decreasing with $B$. Then in the absence of any public or private protection, the probability of being robbed is $1 - F(B/P)$ and expected bandit appropriations are $(1 - F(B/P))Y$, where $Y$ is the total peasant output. This formulation is also similar to Feichtinger and Novak (1994), who assume that bandit loot increases as the peasant income increases. As in Bandiera (2003) and Anderson and Bandiera (2006), we assume that bandit avoidance for peasants is costless and the marginal cost of banditry is zero.

Similar to Anderson and Bandiera (2006) and Konrad and Skaperdas (2012), peasants can be protected by enforcers who prevent some of the thefts. The enforcers are either the public *authority* (feudal lords or the state) or the mafia. We assume that the private protection can be purchased in addition to the existing public protection. The theft success rate is the probability $\pi = (1 - \lambda_A)(1 - \lambda_M)$ where $\lambda_A \in [0,1]$ denotes the control due to public protection and $\lambda_M \in [0,1]$ is the control due to the protection by the mafia. Parameters $\lambda_A$ and $\lambda_M$ measure the enforcement capability of the public authority and the mafia. With protection, bandit appropriations are reduced to

$$R_B = \pi(1 - F(B/P))Y \tag{6.1}$$

We assume that bandits must pay a portion of their loot $R_B$ as a tribute $T_B$ to the mafia. The share of the tribute ranges from zero to some maximum mafia tribute share $\bar{t}_M \in [0,1]$ depending on the mafia strength, which is measured by the variable $\lambda_M$. The bandit tribute share is $\bar{t}_M \lambda_M \in [0,1]$ and bandit payoff to the mafia is $T_B = \bar{t}_M \lambda_M R_B$. Then, bandit disposable income is the loot taken from the peasants less the tribute paid to the mafia: $I_B = R_B - T_B$. The disposable income per bandit is $i_B = I_B/B$. We distinguish between true and perceived incomes. The perceived income per bandit $\hat{i}_B$ is adjusted towards the true income $i_B$ according to the exponential averaging process over period $\tau$:

$$\frac{d\hat{i}_B}{dt} = \frac{i_B - \hat{i}_B}{\tau} \tag{6.2}$$

The payoff for protection services that peasants pay to the mafia is $T_P = p_M M$ where $M$ is the number of *mafiosi* that are protecting the peasants and $p_M$ is the price of protection services. The peasant disposable income $I_P$ is the peasant output $Y$ less bandit appropriations $R_B$ and payoff for services $T_P$: $I_P = Y - R_B - T_P$. The disposable income per peasant is $i_P = I_P/P$. Distinguishing between true and perceived incomes, the perceived income per peasant $\hat{i}_P$ is adjusted according to the exponential averaging process:

$$\frac{d\hat{i}_P}{dt} = \frac{i_P - \hat{i}_P}{\tau} \tag{6.3}$$

where $\tau$ is the perception delay.

Using definitions in Equations 6.2 and 6.3, we specify the economic attractiveness of banditry as $\hat{\imath}_B/(\hat{\imath}_B + \hat{\imath}_P)$, which matches to the interval [0,1]. If the perceived peasant per capita income $\hat{\imath}_P = 0$ and the perceived bandit per capita income $\hat{\imath}_B \neq 0$, then the attractiveness of banditry is one. If $\hat{\imath}_B = 0$, then the economic attractiveness of banditry is zero. We assume that for a non-mafia population $P + B$, the number of potential bandits is $B^* = (P + B)\hat{\imath}_B/(\hat{\imath}_B + \hat{\imath}_P)$. This formulation implies that as economic attractiveness of banditry increases, so does the population who wishes to be bandits. If perceived incomes of peasants and bandits are equal, i.e., $\hat{\imath}_B = \hat{\imath}_P$, then a half of the non-mafia population would like to be bandits and a half would like to continue farming.

The bandit population $B$ changes according to the following derivative:

$$\frac{dB}{dt} = \frac{B^* - B}{\tau_B} \tag{6.4}$$

where $\tau_B$ is the bandit recruitment delay. In this formulation, when economic attractiveness of banditry drops below its previous level, $B^* < B$, the derivative becomes negative, implying that the bandit population $B$ declines because some bandits return to farming.

Following Bandiera (2003), Andersen and Bandiera (2006), and Del Monte and Pennacchio (2012), we assume that peasants' willingness to pay for protection against bandits $W$ is equal to the expected damages from banditry, $W = R_B$. Without loss of generality, we set the peasant budget for protection $L$ to their disposable income, $L = I_P$. Then the spending on protection by peasants is $l = \min(W, L)$.

The price of mafia protection is $p_M = \max(\hat{\imath}_P, \hat{\imath}_B) + c_M$. This formulation implies that the price for *mafiosi* services is the maximum of the perceived peasant and bandit incomes plus some compensating differential $c_M$. Following Del Monte and Pennacchio (2012), the compensating differential $c_M$ can be interpreted as the fixed cost of being a *mafioso*.

The peasant demand for protection is $D_M = l/p_M$. As the demand for protection $D_M$ changes, the population of mafia $M$ adjusts according to the following derivative:

$$\frac{dM}{dt} = \frac{D_M - M}{\tau_M} \tag{6.5}$$

where $\tau_M$ is the mafia recruitment delay. In this formulation, when demand for mafia services declines, i.e., $D_M < M$, the derivative becomes negative and the mafia population $M$ declines.

The strength of mafia is defined by the mafia control variable $\lambda_M = M/m_B$, $\lambda_M \in [0,1]$. Here, $m_B$ is the mafia size required for bandit control. Fewer *mafiosi* are needed to control bandits if there is already some enforcement by the central authority: $m_B = (1 - \lambda_A)\theta_M B$, where $\theta_M \in [0,1]$ is the mafia's marginal protection effectiveness. If, for example, $\theta_M = 0.2$, then one *mafioso* can control five bandits. Parameter $\lambda_A \in [0,1]$ is the authority control.

The state of the economy and society is described by two indices. The first is *lawlessness index*, which is defined as the fraction of the total population engaged in criminal activities: $(B + M)/(P + B + M)$. The index values are in the range $[0,1]$. The second is the *economic integrity index* that denotes the percentage of the total peasant output that is kept by the peasants as their disposable income and not lost to bandits or the mafia. The economic integrity index is defined as $I_P/Y \in [0,1]$.

## 6.6 Feedback analysis

Figure 6.1 shows key variables and important causal and feedback effects as specified by model equations in the previous section. The notation follows the standard graphical notation of system dynamics (e.g. Lane 2000; Maani and Cavana 2007; Morecroft 2007). Rectangle shapes around variables Peasants, Bandits and Mafia indicate that these variables are stocks, or state variables. Bandit recruitment (Equation 6.4) and mafia recruitment (Equation 6.5) are the flows between the stocks. Arrows indicate causal relationships between variables. If an arrow is positive, the relationship between the cause and effect is positive. A negative arrow implies that an increase in one variable causes a decline in the other variable. Circular causal links form reinforcing and balancing feedback loops. A reinforcing loop is shown with the letter R and a balancing loop is shown with the letter B. In this diagram, there are three reinforcing loops and two balancing loops.

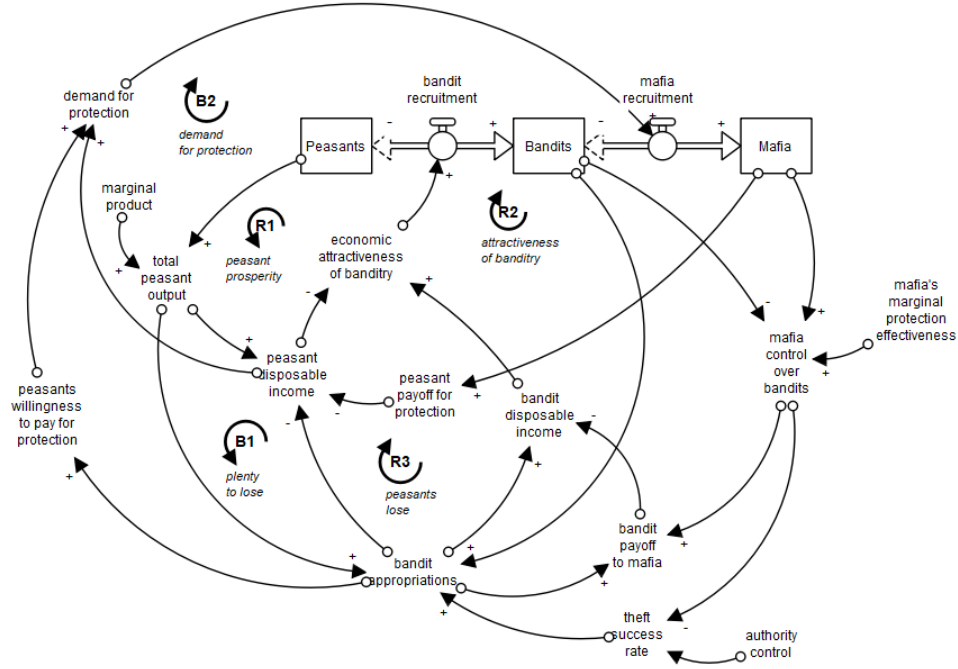

**Fig. 6.1** Key variables and important causal and feedback effects in the model

The model is rich in complex causal relationships and feedback effects. Therefore, a detailed analysis of the model is best performed using specialized system dynamics software such as *Stella Architect* or *Vensim*. Even in its simplified causal form in Figure 6.1, the feedback loop-counting algorithm identified 21 feedback loops passing through the variable Mafia. Each feedback loop that passes through a variable is an influence mechanism. Reinforcing loops attempt to increase the variable, but their influence is weakened by balancing loops. Each loop contains variables that are potential levers for policy intervention. By weakening reinforcing loops and by strengthening balancing loops that pass through undesirable variables, the policymaker can slow down the growth of such variables.

Figure 6.1 shows key economic feedback effects that control the emergence of mafia activity. As the population of peasants increases, they produce more output. Production growth has several effects on banditry that work through reinforcing and balancing loops. First, more output implies that the peasant disposable income rises and, if bandit disposable income stays constant, the relative economic attractiveness of farming as compared to banditry increases, or, in other words, the economic attractiveness of banditry decreases. The declining attractiveness of banditry reverses the flow from the peasants to the bandits. This causal chain forms the reinforcing *peasant prosperity* loop R1. The second effect of the production growth is that as the peasant output grows, so are the possibilities for bandit appropriations (Equation 6.1). Bandit appropriations reduce the peasant disposable income, which means that, if bandit disposable income stays constant, the relative economic attractiveness of farming as compared to banditry decreases, or, in other words, the economic attractiveness of banditry increases. The negative loop *plenty to lose* (B1) balances the reinforcing loop R1.

As bandit appropriations increase, the bandit disposable income increases too, which increases the economic attractiveness of banditry. That encourages more bandit recruitment and leads to more bandit appropriations (reinforcing loop R2, *attractiveness of banditry*). Moreover, as the population of bandits increases, bandit appropriations increase that lowers peasant disposable income, which strengthens the economic attractiveness of banditry. This causal chain forms the reinforcing loop R3 *peasants lose*.

The mounting bandit appropriations contribute to the growth of the peasant willingness to pay for protection and demand for protection. As the demand for protection increases, more *mafiosi* are recruited from the ranks of bandits, which allows the mafia to strengthen its control over bandits and lower the theft success rate. This causal chain completes the balancing loop B2 called *demand for protection* that puts a check on bandit appropriations. Note that the private enforcement by the mafia and public enforcement by an authority are substitutes and both depress the theft success rate.

## 6.7 Scenario experiments

This section describes computational experiments aimed at explaining varying levels of demand for private protection in different scenarios. They show that the model is able to explain empirical facts about the emergence of the Sicilian mafia that have been discussed in previous sections. Because experiments are designed to isolate individual effects, each simulation manipulates only one factor at a time. Experiments that involve changing several parameters at a time are also feasible but they are not explored in this study. The first experiment is the base run. All other experiments are extensions of the base run. The time horizon in the experiments is 25 years, or 300 months.

### 6.7.1 Base run

The objective of this experiment is to show the emergence of mafia activity when the public enforcement is removed. This experiment starts in an equilibrium, which we interpret as the state of the Sicilian economy and society before the land reforms and the unification. Table 6.1 shows equilibrium values. Assume that due to their violent reputation, the mafia requires only 20 percent of the bandits' manpower to control the bandits, i.e. $\theta_M = 0.2$. Marginal product is $a_P = 10$ and it stays constant during the simulation.

**Table 6.1** Parameters and initial values for the base run

| | | |
|---|---|---|
| P | Peasants | 107 |
| B | Bandits | 3 |
| M | Mafia | 0 |
| $a_P$ | marginal product | 10.00 |
| $\theta_M$ | mafia's marginal protection effectiveness | 0.20 |
| $\theta_B$ | effectiveness of bandit technology | 3.00 |
| $c_M$ | fixed cost of being a *mafioso* | 10.00 |
| $\lambda_A$ | authority control | 0.90 |
| $\hat{\imath}_P$ | perceived disposable income per peasant | 9.98 |
| $\hat{\imath}_B$ | perceived disposable income per bandit | 0.28 |
| $\bar{t}_M$ | maximum mafia tribute share | 0.20 |
| $\tau$ | perception delay | 10 |
| $\tau_B$ | bandit recruitment delay | 3 |
| $\tau_M$ | mafia recruitment delay | 5 |

Figure 6.2 shows fractions of the total population that are peasants, bandits and the mafia. The simulation starts with the authority control $\lambda_A$ set to 0.9, which implies strong public enforcement. Because $\lambda_A < 1$, some banditry exists in the initial equilibrium state. In experiments, we will refer to the time when the authority control is reduced to zero as "unification." This is time 60, or $t = 60$, during the simulation. Reducing the authority control to zero at unification, i.e. setting $\lambda_A = 0$, leads to a significant increase in the number of bandits, which drives the demand for private protection and the recruitment of *mafiosi*. After a short period of adjustment, the model reaches a new equilibrium around $t = 110$.

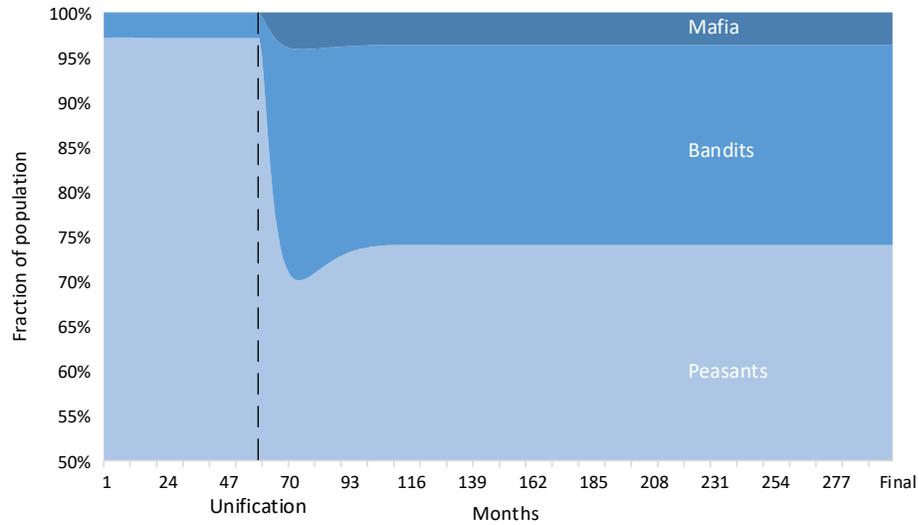

**Fig. 6.2** Relative populations with eliminated authority control

Figure 6.3 provides additional details about the transition to the post-unification equilibrium. Graph (a) shows the evolution of the two indices. The model starts at $t = 0$ and stays for 60 periods in the pre-unification equilibrium characterized by high economic integrity and low lawlessness. Graph (b) shows that during that initial interval, the total peasant output is the highest and mafia is nearly nonexistent. Once the authority control is removed at 60, the economic activity moves into the area of greater lawlessness and lower economic integrity (Figure 6.3a). After the unification, fewer people are involved in productive economic activities, the total peasant output declines, and there is more mafia activity (Figure 6.3b).

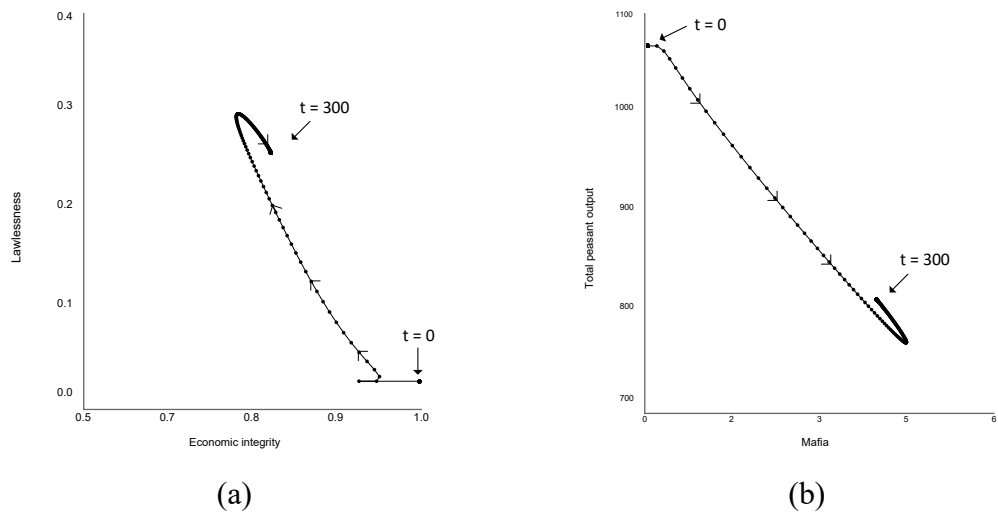

(a)                   (b)

**Fig. 6.3** Emergence of the mafia after authority control is weakened

## 6.7.2 Low output

This experiment explains an empirical observation that banditry was more prevalent and the mafia was less active or not active at all in poorer areas of Sicily (e.g., Del Monte and Pennacchio 2012). The experiment assumes that the marginal product of peasant production $a_P$ is lower than in the base run. We set $a_P = 1$. Lower marginal product leads to lower total peasant output, lower peasant disposable income and greater economic attractiveness of banditry (confirm this causal chain in Figure 6.1). The experiment starts before unification in a low output equilibrium. Table 6.2 shows new equilibrium values that are different from the base run values in Table 6.1.

**Table 6.2** Parameters and initial values for the low output experiment

| | | |
|---|---|---|
| P | Peasants | 100 |
| B | Bandits | 10 |
| M | Mafia | 0 |
| $a_P$ | marginal product | 1.00 |
| $\hat{\imath}_P$ | perceived disposable income per peasant | 1.00 |
| $\hat{\imath}_B$ | perceived disposable income per bandit | 0.10 |

Since public enforcement is strong, i.e. $\lambda_A = 0.9$, only a small fraction of the population is involved in banditry before the unification (Figure 6.4). Weakening authority at time 60 improves the theft success rate $\pi$ and increases bandit appropriations $R_B$, which stimulates the peasant willingness to pay for private protection $W$ (see Figure 6.1). However, due to the low total output, the peasant disposable income $I_P$ is too low for the peasants to afford private protection, which weakens the balancing loop B2. Not constrained by the balancing loop B2, reinforcing loops R2 and R3 drive the increase in banditry (see Figure 6.1). In the absence of enforcement either by the state, landlords or the mafia, the population splits into two classes -- peasants and bandits (Figure 6.4). It should be noted that this model overestimates the attractiveness of banditry and the size of the bandit class since the model does not take into account the social effects that may lower the attractiveness of banditry (see Saeed et al. 2013). This model assumes that peasants consider economic reasons alone when they switch to banditry.

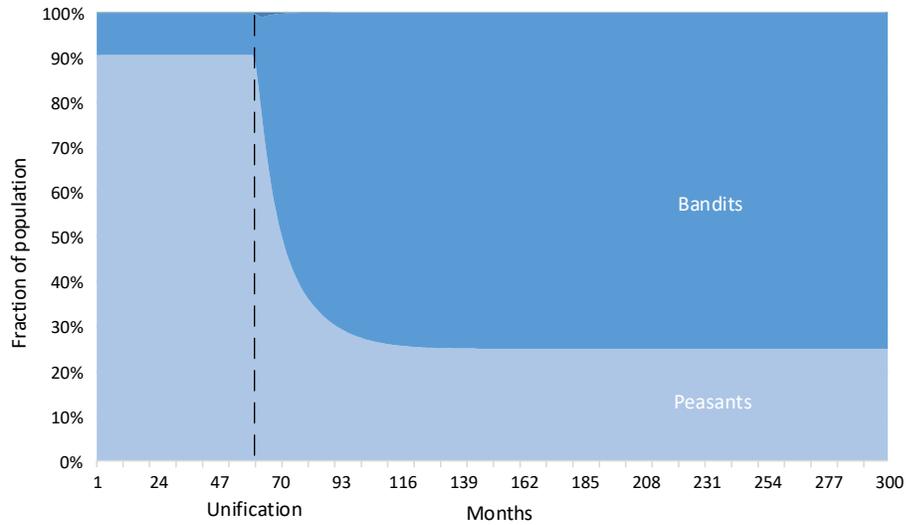

**Fig. 6.4** When output is low, population splits into two groups

The analysis is aided by reviewing the scatter plots in Figure 6.5. At the start of the experiment, $t = 0$, there are relatively few bandits, and therefore peasants keep most of their output, i.e. lawlessness is relatively low and economic integrity is high. By the end of the simulation at $t = 300$, the model reaches an anarchic equilibrium in which there is more lawlessness and less economic integrity (Figure 6.5a). In the new equilibrium, the number of bandits $B$ is high and the total peasant output $Y$ is low (Figure 6.5b).

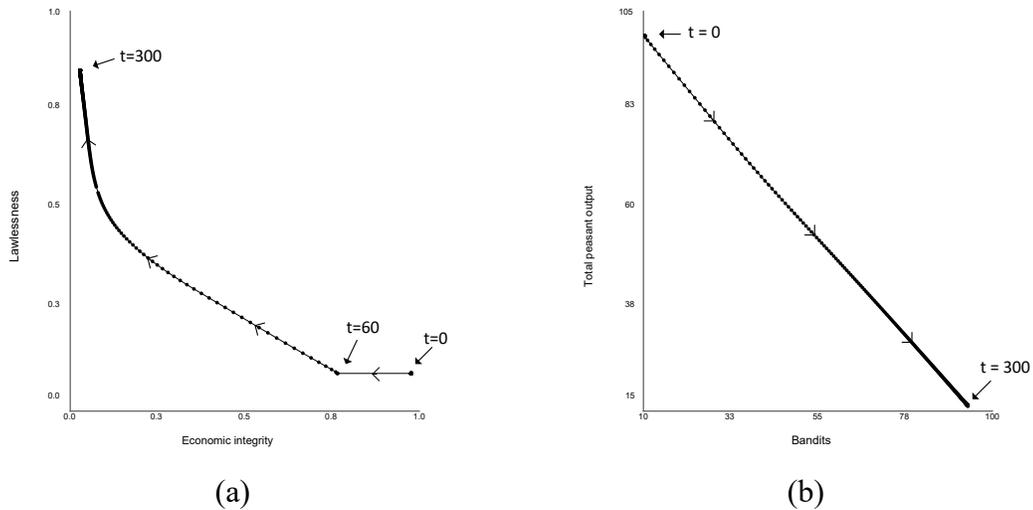

(a)          (b)

**Fig. 6.5** The mafia does not emerge when productivity is low

### *6.7.3 Positive productivity shock*

This experiment tests a hypothesis that a positive productivity shock in low output areas would encourage the emergence of mafia activity. Dimico et al. (2017) argue that an exogenous demand shock for lemons and oranges caused the rise of the Sicilian mafia in citrus-producing regions. Buonanno et al. (2015) document and establish economic causality between the positive demand shock for sulphur and the growth of mafia activity in regions with sulphur production.

This experiment (Figure 6.6) starts as the low productivity run, but we introduce a positive productivity shock at $t = 150$. The trajectories are identical to the trajectories in the low output experiment (Figure 6.4) until 150. At $t = 150$, marginal product is increased from $a_P = 1$ of the low output run to $a_P = 10$. This change increases the total peasant output (confirm causality in Figure 6.1) and strengthens the reinforcing loop R1 that makes peasantry more economically attractive. As bandits abandon their illegal ways, peasant population $P$ increases (Figure 6.6), total peasant output $Y$ increases, which boosts bandit appropriations $R_B$ (see loop B1 in Figure 6.1). With greater willingness to pay for private protection and greater disposable income, peasant demand for protection goes up. That leads to the appearance of the mafia after 150 that imposes control over bandits and reduces economic gains from banditry. This is the balancing loop B2 in Figure 6.1. Lower bandit appropriations due to mafia control convince more bandits to return to productive activities.

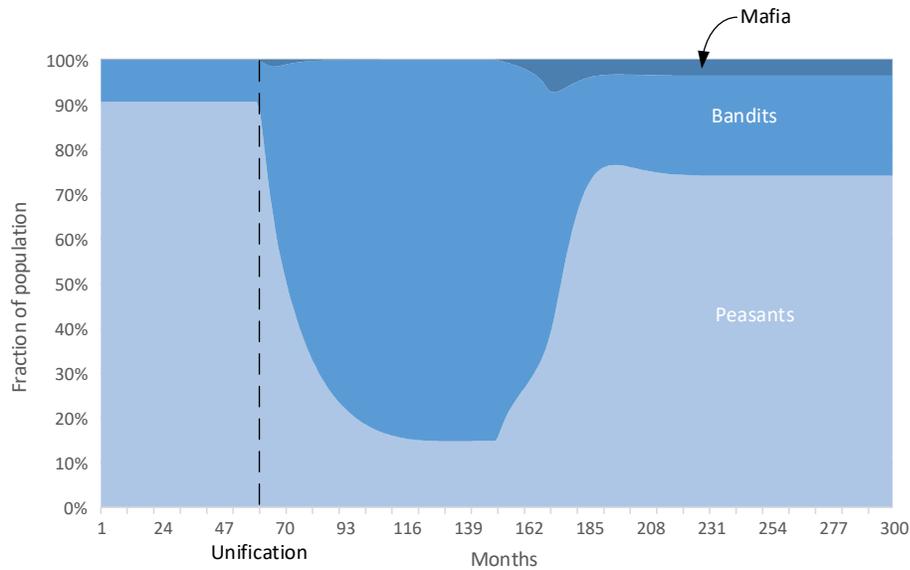

**Fig. 6.6** Productivity shock at 150 leads to the emergence of the mafia

Figure 6.7a tracks economic integrity and lawlessness for this experiment. After the "unification" at $t = 60$, the model converges to the same anarchic equilibrium as in the low output experiment earlier when economic integrity is low and lawlessness is high. However, after the productivity shock at 150, the model moves to the new mafia-

controlled equilibrium with higher economic integrity and reduced lawlessness. Figure 6.7b confirms that the total peasant output $Y$ increases when a stronger mafia presence brings more control over the bandits.

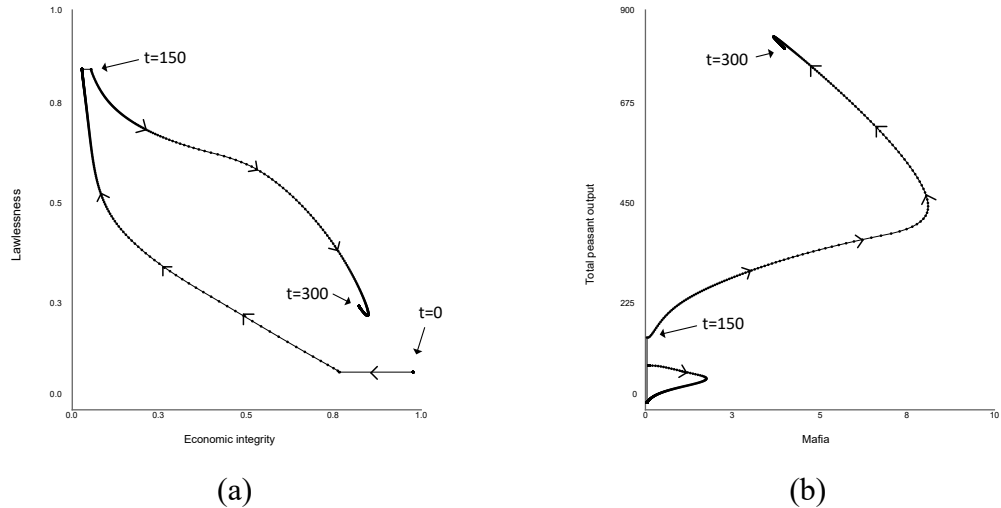

(a)                  (b)

**Fig. 6.7** Emergence of the mafia reduces banditry and lawlessness

### *6.7.4 Eliminating the mafia*

Buchanan (1980) observed that organized crime might be beneficial for the economy. Indeed, in Sicily, banditry was less prevalent in areas with higher levels of organized crime (Del Monte and Pennacchio 2012), which also happened to have higher levels of output (Buonanno et al. 2015; Dimico et al. 2017). This experiment tests the consequences of eliminating the mafia.

We start with the base run. As before, after public enforcement is removed post-unification, i.e. $\lambda_A = 0$ after 60, the model moves to a mafia-enforced equilibrium (Figure 6.8). Now consider that at $t = 150$ the *demand for protection* loop B2 is eliminated and the demand for private protection becomes $D_M = 0$. Removing the mafia creates a lack of enforcement as mafia control over bandits $\lambda_M$ drops to zero. Because reinforcing loops R2 and R3 (Figure 6.1) are no longer constrained by the mafia, the model moves to an insecure equilibrium, in which peasants are dominated by predating bandits.

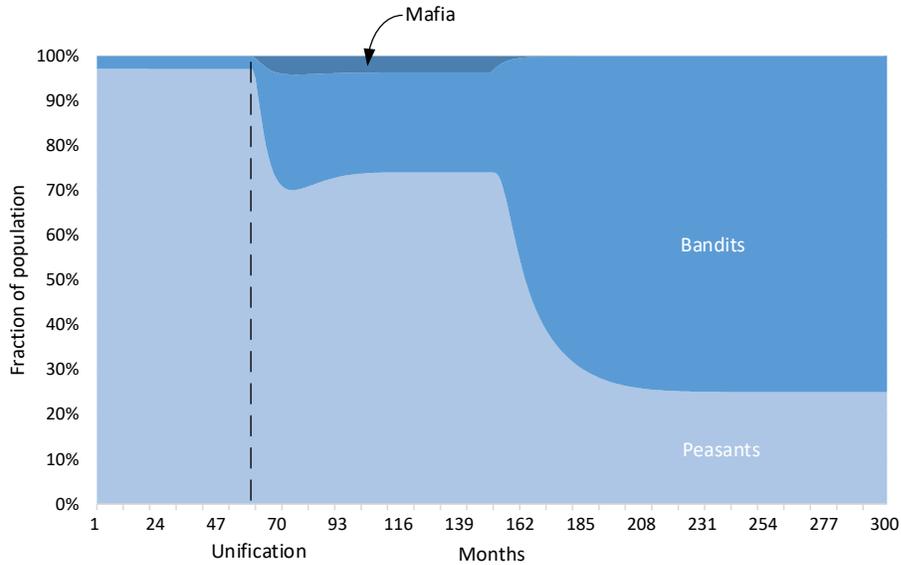

**Fig. 6.8** Banditry is rampant without the mafia to control it

Figure 6.9 provides additional details. After the authority control $\lambda_A$ is reduced at the time of "unification," banditry increases leading to the equilibrium characterized by more lawlessness and lower economic integrity (Figure 6.9a). However, the situation worsens further after mafia activity is eliminated starting at 150. Figure 6.9b shows the total peasant output $Y$ with and without the mafia. At time $t = 0$, when authority control is strong, there is no mafia activity and the total peasant output is high. After the "unification" at period 60, the situation changes. By $t = 150$, the mafia is present and the total peasant output is lower than at $t = 0$. However, without the protection of the mafia after $t = 150$, banditry attracts a big fraction of the population (Figure 6.8) and the total peasant output drops to its lowest level (Figure 6.9b).

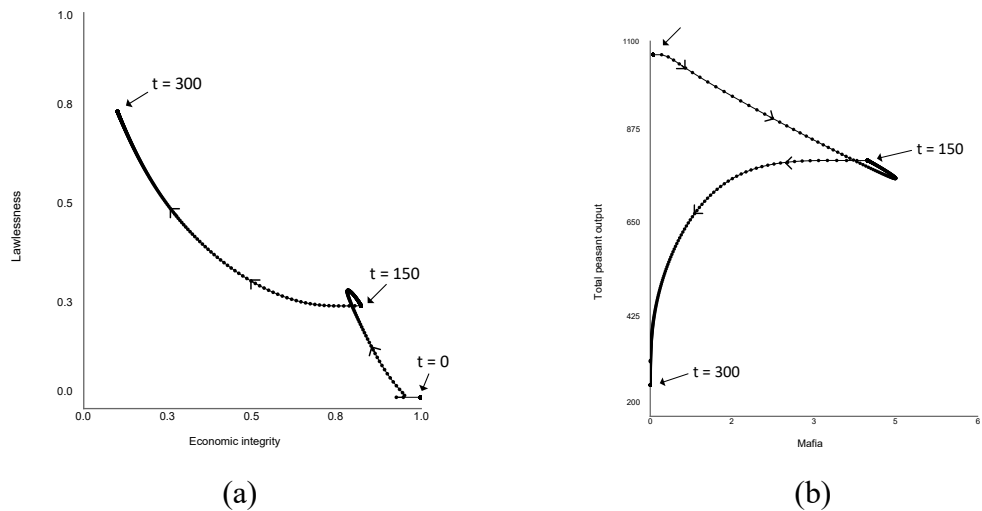

(a)        (b)

**Fig. 6.9** In an anarchic economy without the mafia, output drops

## *6.7.5 No bandits*

This experiment examines a situation when through collective action peasants overcome the free-rider problem (Olson 1965) by eradicating banditry. Elimination of private property rights in favor of communal property may potentially lead to this outcome.

We start with the base run, which removes authority control at 60 (Figure 6.10). Increased success of theft encourages predatory behavior and, as before, the economy settles into the mafia-controlled equilibrium. However, at $t = 150$, we eliminate banditry by setting the number of potential bandits to zero, $B^* = 0$. Very quickly, current bandits go back to productive activities. Since there is no threat from bandits and bandit appropriations $R_B$ drop to zero, the peasants are no longer willing to pay for protection, and therefore mafia demand and the mafia disappear (see Figure 6.1).

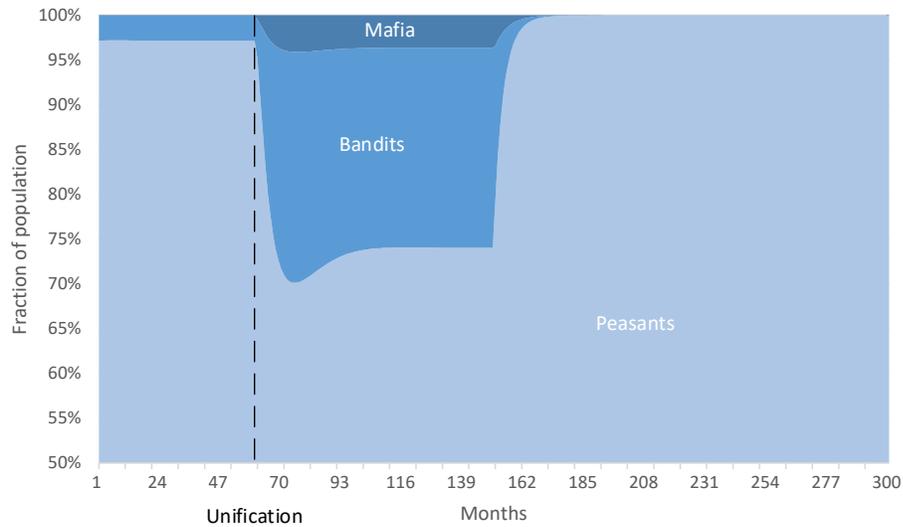

**Fig. 6.10** Solving the banditry problem eliminates the demand for mafia

By the end of the simulation at $t = 300$, the model settles into a new equilibrium that has no bandits or mafia, which implies that the lawlessness index is zero and economic integrity is one (Figure 6.11a). When there are no bandits and no mafia activity, the total peasant output $Y$ is at its highest level (Figure 6.11b).

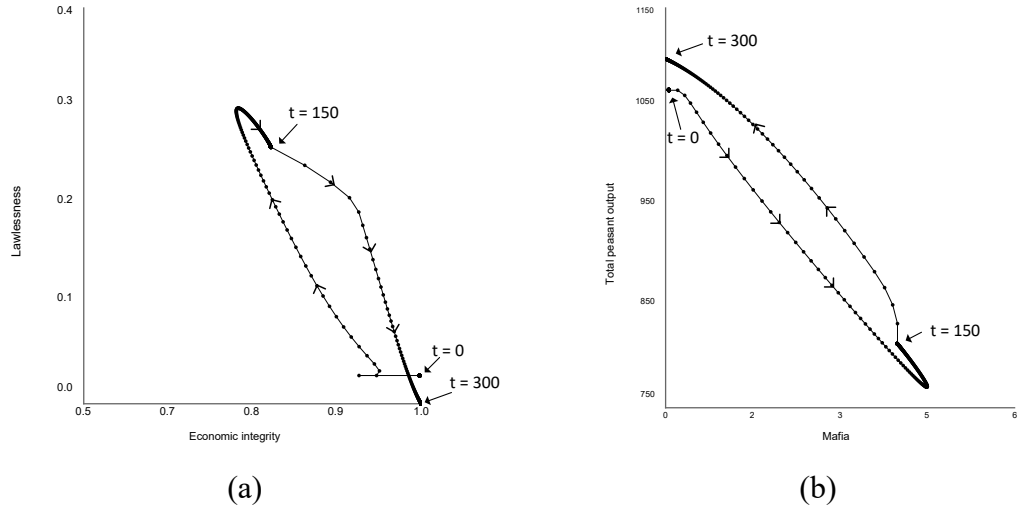

(a)                          (b)

**Fig. 6.11** The economy is better without the bandits

### *6.7.6 Increasing state control*

Under Fascism in the 1920s and 1930s, the mafia as a parallel enforcement structure was substituted by representatives of the Fascist party (Blok 1969; Gambetta 1993). This experiment simulates the effect of a stronger state on the mafia.

    We start with the base run, in which the model settles to the mafia-controlled equilibrium after the "unification" (Figure 6.12). But then as the state become stronger, the authority control is increased at $t = 150$ from $\lambda_A = 0$ to $\lambda_A = 0.5$. Notice that the experiment sets the authority control to the level below the authority control prior to the "unification," which was $\lambda_A = 0.9$. Increasing the state enforcement reduces the theft success rate $\pi$, which has two effects: it reduces bandit appropriations $R_B$ and it increases peasant disposable income $I_B$ (confirm in Figure 6.1). Through reinforcing loops R1, R2 and R3, both effects reduce the economic attractiveness of banditry leading to fewer bandits (Figure 6.12). Additionally, since public and private enforcements are substitutes, a stronger state weakens the balancing loop B2 (Figure 6.1) by cutting into the demand for mafia protection services, $D_M$. Consequently, the number of *mafiosi M* declines. Increasing state control results in a steady-state equilibrium with fewer mafia and a greater fraction of the population in productive activities.

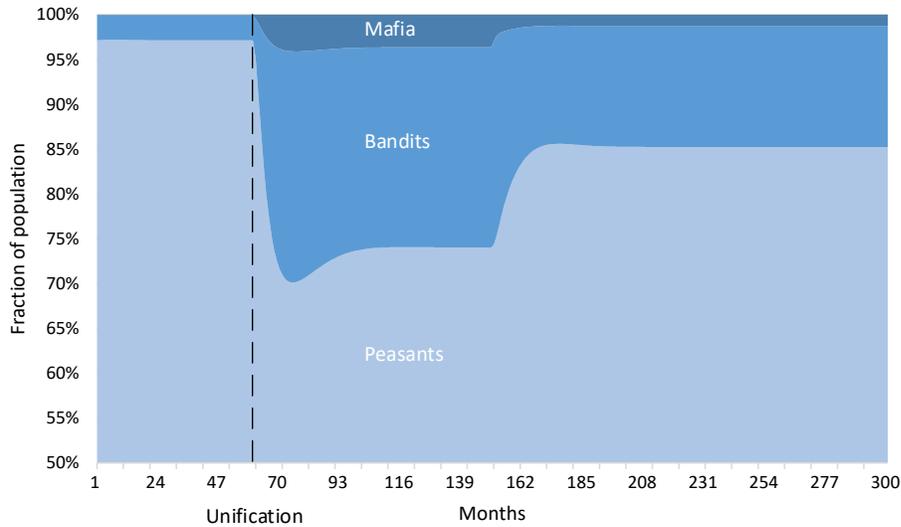

**Fig. 6.12** Increased state control reduces banditry and the need for private protection

Strengthening the state at $t = 150$ proves to be an effective way to decrease lawlessness and increase the amount of output retained by the peasants, which boosts economic integrity (Figure 6.13a). Additionally, more state control has a positive effect on the total peasant output, while decreasing the mafia influence (Figure 6.13b).

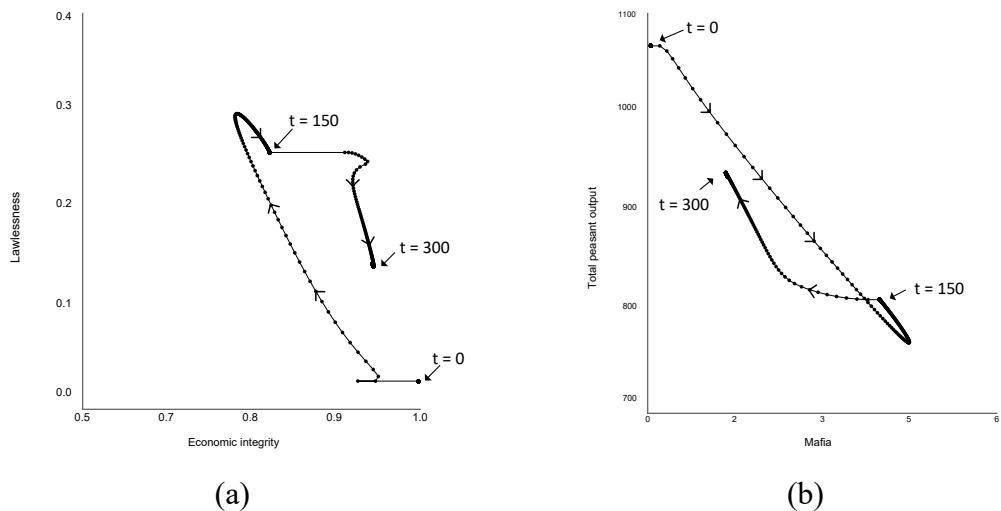

(a)          (b)

**Fig. 6.13** A stronger state leads to better economy

### *6.7.7 Summary of experiment outcomes*

Table 6.3 provides a summary of experiment outcomes. The *base run* and the *low output* experiment show that a weakened law enforcement leads to more banditry. Moreover, the mafia does not emerge in the *low output* experiment, which corresponds to the empirical observation that poor regions were less likely to see active mafia organizations (e.g., Del

Monte and Pennacchio 2012). The *positive productivity shock* simulation demonstrates that if a low output region experiences a productivity shock, the mafia may step in to rein in the bandits, which pushes down the lawlessness index.

The last three experiments have explored scenarios with less organized crime. In the experiment *eliminating the mafia*, economic integrity drops and lawlessness increase because the bandit population swells as it is no longer constrained by the mafia. Overcoming the banditry problem as in the *no bandits* experiment, eliminates the demand for private protection services by the mafia. Therefore, the lawlessness index drops and economic integrity surges. The experiment *increasing state control* simulates stronger law and order. With stronger public enforcement, illegal activities by the bandits drop, and therefore the demand for the mafia's services declines too. Because the peasants' output and disposable income increase, the economic integrity index improves.

Table 6.3 Experiment outcomes

| Experiments | Peasants | Bandits | Mafia | Lawlessness | Economic integrity |
|---|---|---|---|---|---|
| Base run | - | + | + | + | - |
| Low output | - | + | n/c | + | - |
| Positive productivity shock | + | - | + | - | + |
| Eliminating the mafia | - | + | - | + | - |
| No bandits | + | - | - | - | + |
| Increasing state control | + | - | - | - | + |

+ value increased
- value decreased
n/c no change in value

## 6.8 Conclusion and future research

We have developed a *simulation feedback model* that explains the emergence of the mafia as a deterrent of predation on producers in the environment of weak law and order. The model identifies key causal relationships and feedback effects that exist due to the interaction of honest peasants, predatory bandits and the mafia who sell protection services. It explains empirical observations about the Sicilian mafia including the cross-regional differences in mafia activity. The model has also been used to test scenarios involving the elimination of the mafia and the banditry.

By focusing on economic causes only, this study does not present a complete picture as it does not account for factors beyond economics (Allum and Sands 2004). However, the model can serve as a starting point for future research. In this version of the model, the mafia discourages predation by making banditry less attractive economically, rather than through threats of violence. Of course, violence has always been an important enforcement mechanism. Therefore, a future model can consider direct and indirect effects of violence. It would likely show a strong feedback loop that reinforced the growing strength of the mafia as more people asked for protection and its reputation for violence grew.

Future research can provide insights into the tolerance of the ruling elites towards the mafia as the source of the mafia's influence. The legitimacy of the mafia came from the services it provided to the upper classes (Bandiera 2003). For example, Acemoglu et al. (2020) show that the need of the ruling elites to suppress the Fasci peasant movement following the severe droughts of 1893, triggered further proliferation of the mafia. Additionally, the model can endogenize the impact of protection enjoyed by the mafia when organized crime infiltrates executive, judicial and legislative branches of the government (Gambetta 1993).

Because enforcers of security and protection can easily turn to rent seeking (Konrad and Skaperdas 2012; Dimico et al. 2017), the mafia's protection and extortion businesses are hard to differentiate (Buonanno et al. 2015). Historically, the mafia has been involved in extortion and many other illegal activities such as the drug trade (Gambetta 1993). Including extortion in the model is likely to provide insights into the negative impact of the mafia on long-term economic outcomes.